\documentclass{elsart}

\usepackage{graphics}
\usepackage{epsfig}
\usepackage{amsmath}
\usepackage{amssymb}
\usepackage{multirow}
\usepackage{rotating}

\begin{document}

\begin{frontmatter}

\title{Value-at-Risk and Tsallis statistics: risk analysis of the aerospace sector}

\author{Adriana P. Mattedi$^1$, Fernando M. Ramos$^1$, Reinaldo R.
Rosa$^1$ }
\author{Rosario N. Mantegna$^2$ }

\address{$^1$ N\'ucleo para Simula\c{c}\~ao e An\'alise de Sistemas
Complexos - Laborat\'orio Associado de Computa\c{c}\~ao e
Matem\'atica Aplicada - Instituto Nacional de Pesquisas Espaciais
(INPE) - S\~ao Jos\'e dos Campos, Brazil}

\address{$^2$ Istituto Nazionale per la Fisica della Materia, Unit\`a di
Palermo and Dipartimento di Fisica e Tecnologie Relative,
Universit\`a di Palermo, Viale delle Scienze, I-90128, Palermo,
Italia}

\begin{abstract}
In this study, we analyze the aerospace stocks prices in order to
characterize the sector behavior. The data analyzed cover the
period from January 1987 to April 1999. We present a new index
for the aerospace sector and we investigate the statistical
characteristics of this index. Our results show that this index
is well described by Tsallis distribution. We explore this result
and modify the standard Value-at-Risk (VaR), financial risk
assessment methodology in order to reflect an asset which obeys
Tsallis non-extensive statistics.
\end{abstract}

\end{frontmatter}

\section{Introduction}

Financial institutions are subject to many sources of risk.
Financial risk can be broadly defined as the degree of
uncertainty about future net returns. Accurate evaluation of
risks in financial markets is of crucial importance for the
proper assessment and efficient mitigation of risk and,
consequently, for optimal capital allocation. The increased
volatility of financial markets during the last decade has
induced the development of sophisticated risk management tools.
In this context, Value-at-Risk (VaR) has become the standard
measure that financial analysts use to quantify market risk. VaR
is defined as the maximum potential loss in a portfolio value due
to adverse market movements, for a given probability (5\% or 1\%,
for example) and a fixed time horizon (typically, 1 day). The
great popularity achieved by this instrument is essentially due
to its conceptual simplicity: VaR reduces the market risk
associated with any portfolio to just one number, the loss
associated to a given probability, under {\it normal} market
conditions.

Here, we analyze the financial risk of the aerospace sector. The
aerospace complex "weights" several hundreds of billions of
dollars a year, and creates millions of high-wage, high skill
jobs worldwide. Moreover, technologically-intensive aerospace
production generates extremely important positive spill over
effects for the rest of the economy. However, this very fact --
the need of large investments in advanced technologies -- makes
this industry highly susceptible to political and economic risks.
For example, if we consider the volatility as a risk measure, we
remark that, in the last decade, the stocks of the main aerospace
players presented a volatility 20\% higher than that displayed by
major financial indexes such as the Dow-Jones Industrial Index and
the Standard \& Poor's 500 Index (see below). Since until
presently, there is no single market indicator specially designed
for tracking the performance of the aerospace sector as a whole,
we introduce a new index, called Comprehensive AeroSpace Index
(CASI), based on the stocks negotiated on the New York Exchange
(NYSE) and the Over the Counter (OTC) markets. We investigate the
statistical characteristics of the CASI and show that this index
is well described by Tsallis distribution. We explore this result
and modify the standard Value-at-Risk (VaR) risk assessment
methodology in order to reflect an asset whose underlying
dynamics obeys the non-extensive statistics.

This paper is organized as follows. Following this Introduction,
in Section 2 we describe the data set used in this work, how the index CASI is computed, and
which are its main statistical
characteristics. In Section 3, we briefly introduce the VaR methodology,
then generalize it to non normal, Tsallis market conditions, and
apply this new methodology to the CASI time-series. Finally, our concluding remarks are
presented in Section 4.

\section{The Comprehensive Aerospace Index (CASI)}

The data set used in this paper consists of twelve years of daily
prices sets, from January, 1987 to April, 1999, from a
comprehensive sample of leading aerospace-related companies with
stocks negotiated on the New York Exchange (NYSE) and on the
Over-the-Counter (OTC) markets. The number of aerospace companies
included in the data set varied from 62 in 1987, to 146 in 1999.
The CASI was calculated using a base-weighted aggregate
methodology \cite{Nysemet,SPmet} in order to reflect the total
market value of each company. Following this approach, the CASI
at a given time $t$ is given by
\begin{equation}
\label{he1}
Y(t) = \frac{1}{ID} \, \sum_i^N (\frac {w_i}{W}) \, X_i (t),
\end{equation}

where $w_i$ is the number of shares outstanding of the stock $i$,
$W$ is the total number of companies shares that compose the
index, $X_i$ is the price of the stock $i$, and $N$ is the number
of stocks. ID is a normalizing factor called Index Divisor, which
permits to set the initial CASI value to an arbitrary level (say,
100 or 10), allowing the index series to be comparable over time,
and to adjust for variations in the weights $w_i$. These weights
are up-dated every three months, or every time a difference of
5\% or more in the number of shares outstanding of any stocks is
detected or when a new stock is included in the index. In any
case, the ID value is correspondingly adjusted to assure the
continuity in the CASI index values.

\begin{figure}[ttt]
\begin{center}
    \mbox{\epsfig{file=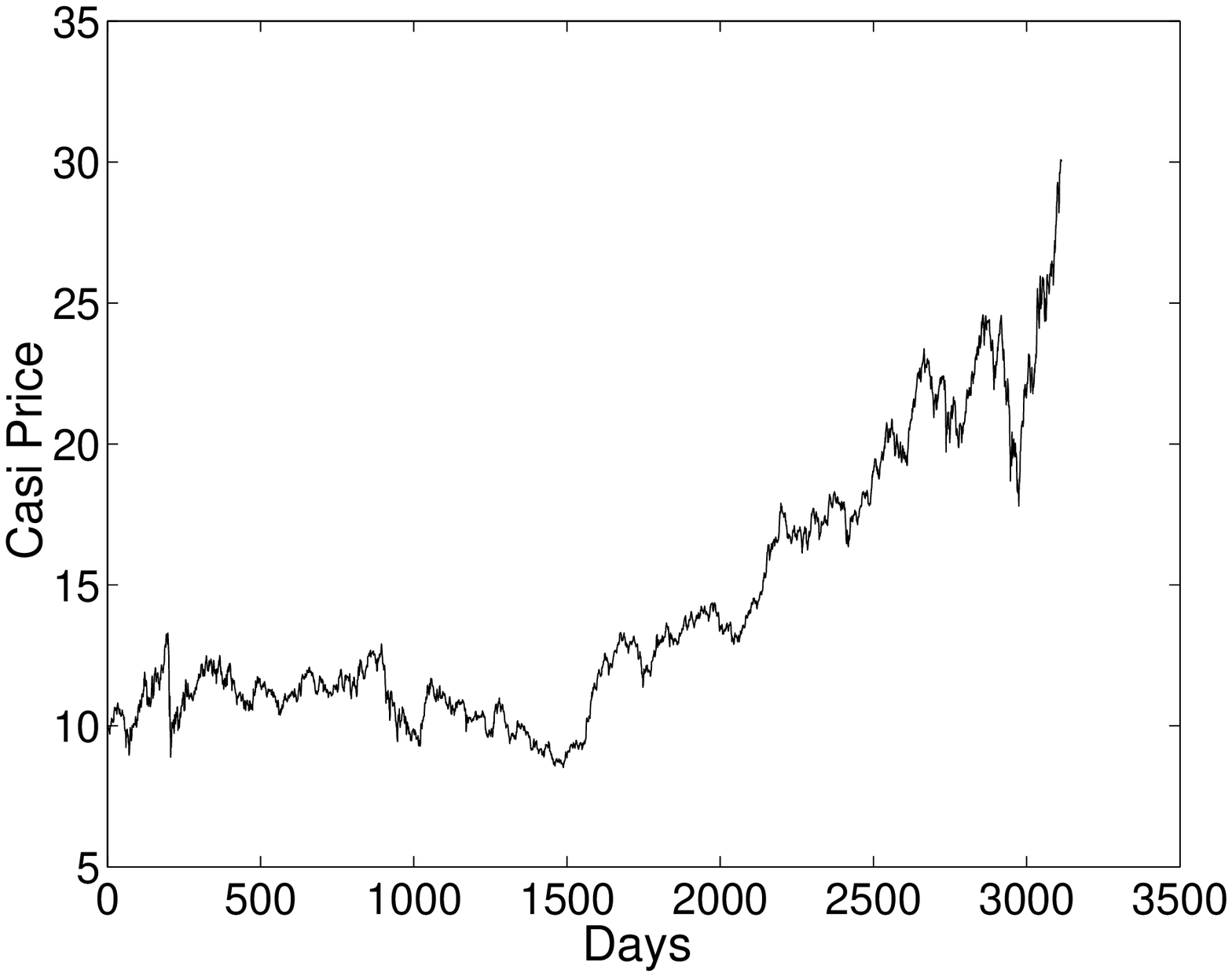,width=7cm,height=5cm}}
    \mbox{\epsfig{file=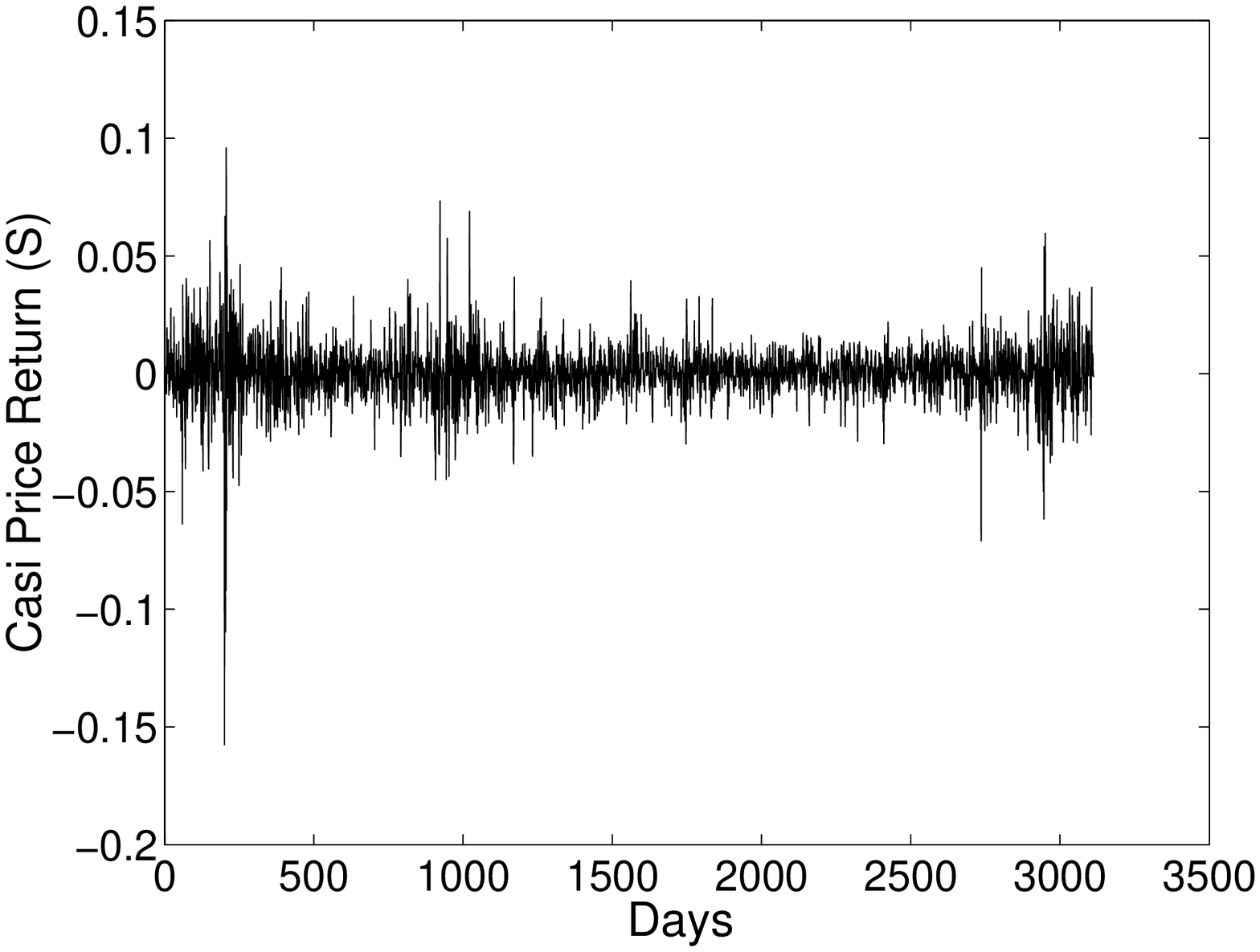,width=6.5cm,height=5cm}}
    \caption{The CASI index: (a) daily evolution; (b) daily returns.}
    \label{indice}
\end{center}
\end{figure}

\begin{figure}[ttt]
\begin{center}
    \mbox{\epsfig{file=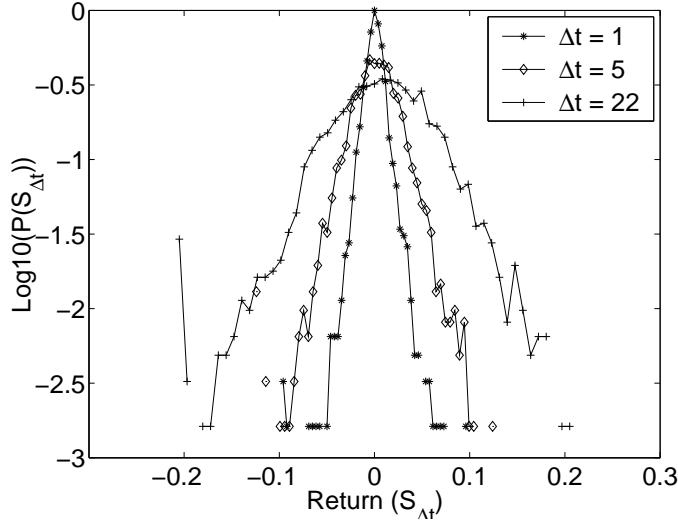,width=9cm,height=7cm}}
    \caption{CASI return histograms, for $\Delta t$ (1, 5, 22 days).}
    \label{casi_pdf}
\end{center}
\end{figure}

As usual, we considered in our analysis the time evolution of
CASI returns, defined as the logarithmic differences of CASI
values over a period of time $\Delta t$, computed for a sequence
of index values as $S_{\Delta t}(t) = \log(Y(t + \Delta t)/Y(t))$.
CASI returns, from 1987 to 1999, are shown in Fig. \ref{indice},
for a time delay of 1 day, together with the time evolution of
the index itself during the same period. The spectral
characteristics of both signals display some well-known features
of financial time-series, with CASI returns behaving like a
white-noise and the CASI index itself following a random-walk
pattern. We present in Fig. \ref{casi_pdf}, CASI return
histograms (in semi-logarithmic plots), for three time intervals.

These results follow a familiar pattern already found in other
financial signals \cite{MantStanN,Ghashghaie,RamosNA,Mattedi}. The
transition from an approximately Gaussian behavior, at long time
intervals, to a leptokurtotic distribution form, as $\Delta t$
decreases, is quite evident. At $\Delta t = 1$ day, the histogram
has a kurtosis considerably higher than 3, the value expected for
a normal process. This feature, which corresponds to an excess of
large fluctuations compared to a Gaussian distribution is a well
known signature of intermittence - bursts of intense market
volatility.

\begin{table}[bbb]
\begin{center}
\begin{tabular}{ccccc}
            \multicolumn{5} {c} {Table 1: Scaling coefficients} \\
            \multicolumn{5} {c} {} \\ \hline
           & \multicolumn{4} {c} {Indexes} \\ \cline{1-5}
           & CASI    & S\&P500  & DJI     & Nasdaq    \\ \hline
 Mean      & 0.96  & 0.99  & 0.96  & 0.98    \\ \hline
 Variance  & 1.00  & 0.93  & 0.96  & 1.09    \\ \hline
 Kurtosis  & -0.38 & -0.58 & -0.61 & -0.32   \\ \hline
\end{tabular}
\end{center}
\end{table}

\begin{figure}[ttt]
\begin{center}
    \mbox{\epsfig{file=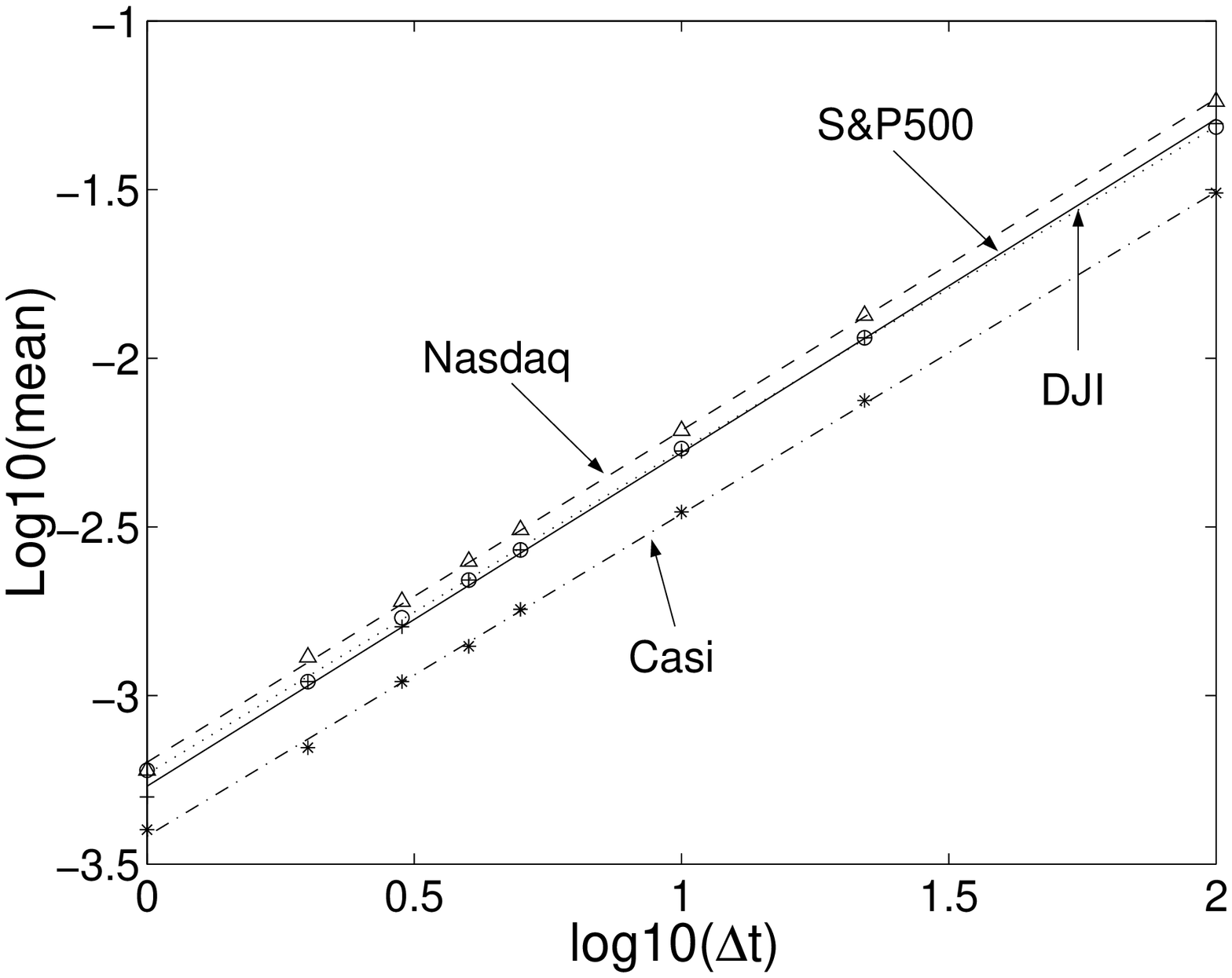,width=8cm,height=6.5cm}}
    \mbox{\epsfig{file=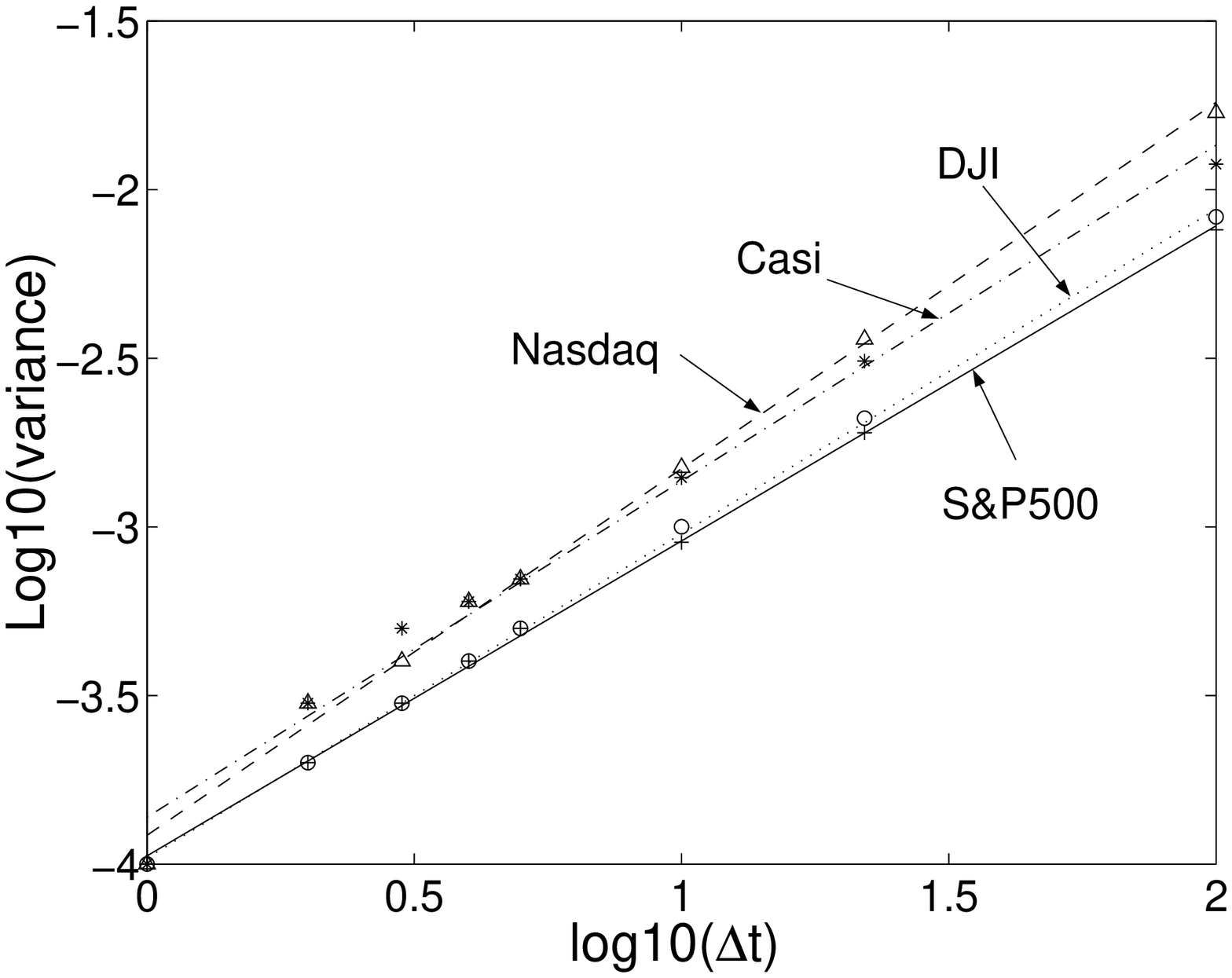,width=8cm,height=6.5cm}}
    \mbox{\epsfig{file=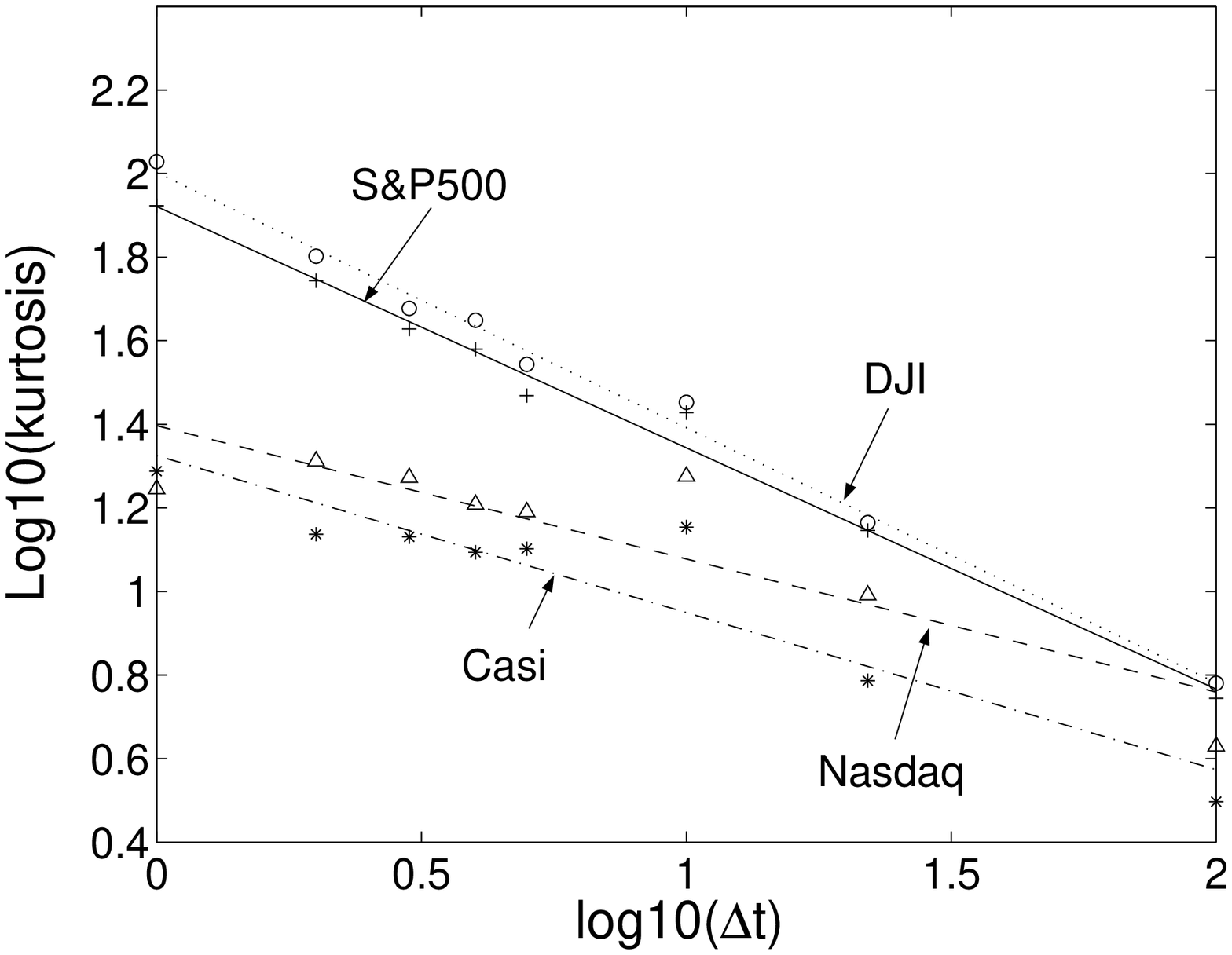,width=8cm,height=6.5cm}}
    \caption{Scaling with varying time intervals
of CASI, Nasdaq, DJI and S\&P500: (a) mean; (b) variance; (c)
kurtosis.}
    \label{scaling}
\end{center}
\end{figure}

Financial markets are extremely complex, with multiple
hierarchical structures and a large number of heterogeneous
agents interacting in an intricate way. When studying the
statistical characteristics of a financial index, an usual
approach is to investigate how the underlying distribution moments
(or other related quantities) scale for different time intervals.
In Fig. \ref{scaling} we present, on a log-log plot, mean,
variance and kurtosis as a function of $\Delta t$, for four
financial indexes: CASI, Nasdaq, Dow-Jones Industrial (DJI) and
Standard \& Poor's (S\&P500). We remark that all curves present a
power-law behavior (the corresponding scaling exponents are
displayed in Table 1), although the kurtosis results suffer from
the finiteness of the sample size. For all indexes, the variance
scales linearly with $\Delta t$, the expected result for a
standard diffusive regime. Note that CASI and Nasdaq have
kurtosis exponents quite different from those displayed by DJI
and S\&P500. We may speculate that this result indicates that the
dynamics of price formation in technology-intensive companies is
peculiar, very different from more traditional industries.
Surprisingly, we find that, although CASI and Nasdaq consistently
display higher variances, their kurtosis are smaller. In other
words, CASI and Nasdaq are more volatile but less intermittent
than DJI and S\&P500. This somewhat unexpected result suggests
the presence of long memory effects in CASI and Nasdaq volatility
patterns. These correlations tend to cluster periods of similar
volatility, preventing the excessive alternating of price changes
regimes.

\section{Risk Analysis}

In recent years, VaR has been adopted as the standard tool used by
institutions to measure and manage financial market risk.
Moreover, VaR is the basis for BIS (Bank for International
Settlements) market risk-based capital requirements
\cite{Jorion,Saunders}. Although, its origins can be trace back
as far as 1922, to capital requirements the NYSE (New York Stock
Exchange) imposed to member firms, passing through the portfolio
theory and a first crude VaR measure published in 1963
\cite{Holton}, VaR only became widely used with the J.P.Morgan's
RiskMetrics Technical Document in 1995 (freely available on the
J.P.Morgan's website \cite{Morgan}). Great part of its popularity
originates from the fact it aggregates several risk components
into a single number. VaR measures the worst expected loss under
{\it {normal}} market conditions over a specific time interval
(which may vary from hours to years), at a given confidence level
(for example, 99\% or 95\%).

In general terms, VaR measures the quantile of the projected
distribution of gains and losses over a given time horizon. The
$q$-quantile (or fractile) of a random variable $X$ is any value
$x$ such that $Pr(X \leq x) = q$. If $c$ is the selected
confidence level, VaR is the corresponding $1-c$ lower-tail
quantile. Analytically, it can be formulated in terms of returns
of the portfolio as follows:
\begin{equation}
\label{varmean}
Pr[S_{\Delta t} < VaR_S(\Delta t)] = 1 - c~,
\end{equation}
where $VaR_S(\Delta t)$ is the VaR, in terms of returns, for a
time horizon of $\Delta t$.

It is clear from Eq. (\ref{varmean}), that the precise prediction
of the probability of an extreme movement in the value of a
portfolio is essential for risk management. By their very nature,
extreme movements are related to the tails of the distribution of
the underlying data generating process. Several studies, starting
with the pioneering work by Mandelbrot \cite{Mandelbrot}, have
shown that financial return distributions are leptokurtotic, that
is they have heavier tails and a higher peak than a normal
distribution. These features are well illustrated by CASI 1-day
return distribution, shown in Fig. \ref{casi_pdf}. In practical
terms, the assumption of normally distributed returns may lead to
serious errors in the estimation of risk, affecting the stability
of markets and/or the profitability of financial institutions.

In order to model fat-tailed distributions, the log-normal
distribution, generalized error distribution, truncated Levy
distribution, and mixtures of normal distributions have been
suggested \cite{MantStanlivro,Ramoscondmat,Michael,Ausloos,Liu}.
From an econophysics point-of-view, several studies pointed out
that the financial market dynamics may efficiently be understood
within the statistical mechanical framework and its suitable
generalizations \cite{Michael,Ausloos}. In this context, Ramos
and collaborators \cite{Ramoscondmat,RamosNA} have recently shown
that Tsallis canonical distribution \cite{Tsallis}, given by
\begin{equation}
\label{he10}
p_q(x) \sim (1 - \beta(1 - q)\, x^2\,)^{1/(1 -
q)}~,
\end{equation}

with $q$ and $\beta$ as parameters, provides a simple and accurate
model for financial histograms. Moreover, Tsallis entropic
parameter $q$ was found to be an excellent measure of
intermittency \cite{Bolzan}. Note that, for $1 < q < 5/3$, Eq.
(\ref{he10}) displays fat-tails but still has finite variance.
For $q=1$, we recover the Gaussian distribution. Finally, for
unit variance, we have $\beta = 1/(5-3q)$.

Relaxing the assumption of normal returns intrinsic to the
original VaR method, and assuming that market return
distributions follow Eq. (\ref{he10}), $VaR_S(\Delta t)$ can be
computed from

\begin{equation}
\label{he12}
Z_q^{-1} \, \int_{-\infty}^{-VaR_S(\Delta t)}
[1 - \beta(1 - q)S^2]^{1/(1 - q)}dS = 1-c~,
\end{equation}
where $q$ is a function of $\Delta t$ and, for unit variance,
$\beta = 1/(5-3q)$. The normalizing factor is given by
\begin{equation}
Z_q = \sqrt{\frac{\pi}{\beta (q-1)}} \,
\frac{\Gamma((3-q)/2(q-1))}{\Gamma(1/(q-1))}~.
\end{equation}
It is important to remark that, since the entropic parameter $q$
only depends on the kurtosis of the experimental histogram
\cite{Bolzan}, its value can be easily estimated for longer time
horizons. For this, it suffices to extrapolate the kurtosis value
of the 1-day histogram, assuming some known scaling with $\Delta
t$, as suggested by Fig. \ref{scaling}c.

\begin{figure}[ttt]
\begin{center}
    \mbox{\epsfig{file=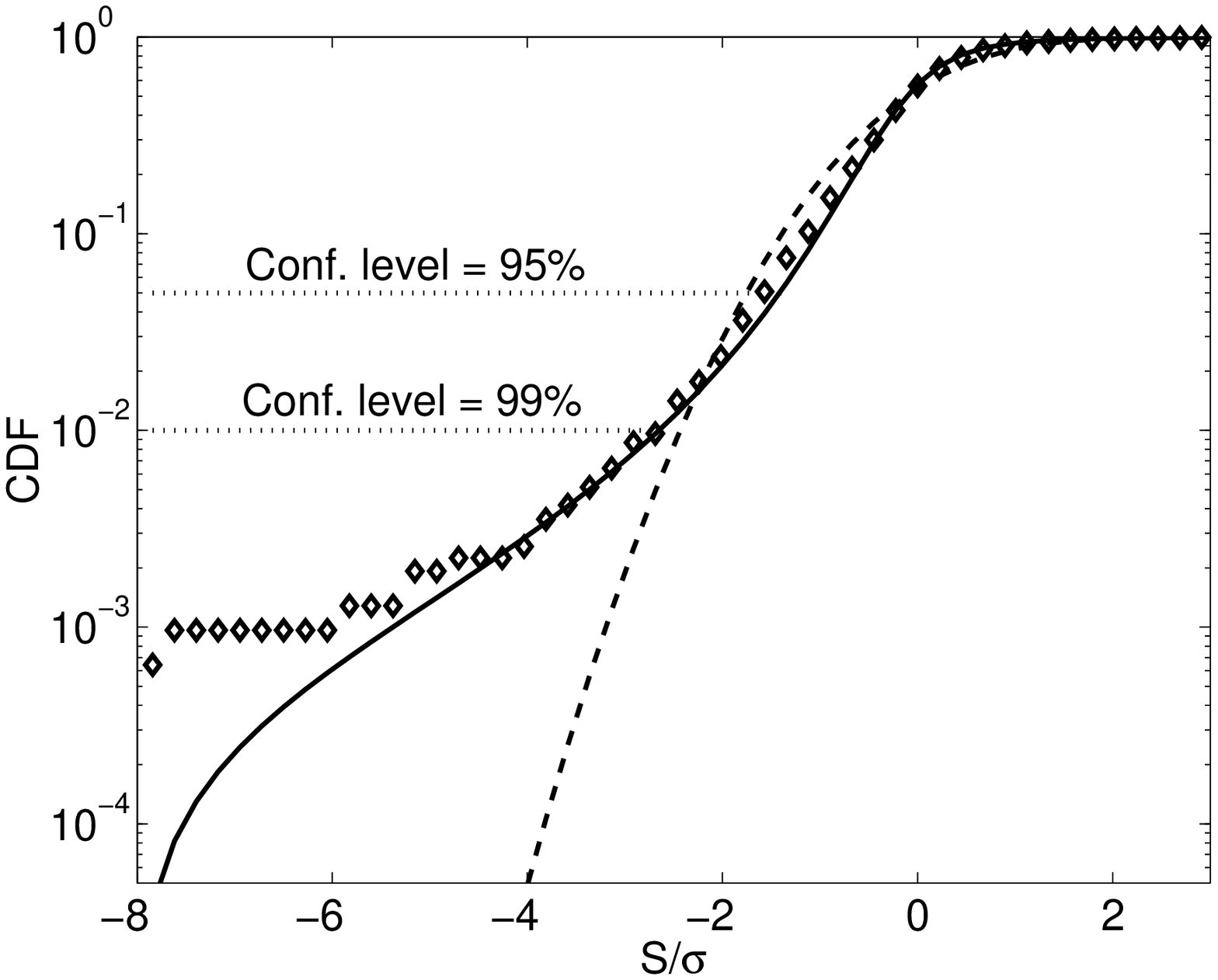,width=6.7cm,height=6.2cm}}
    \mbox{\epsfig{file=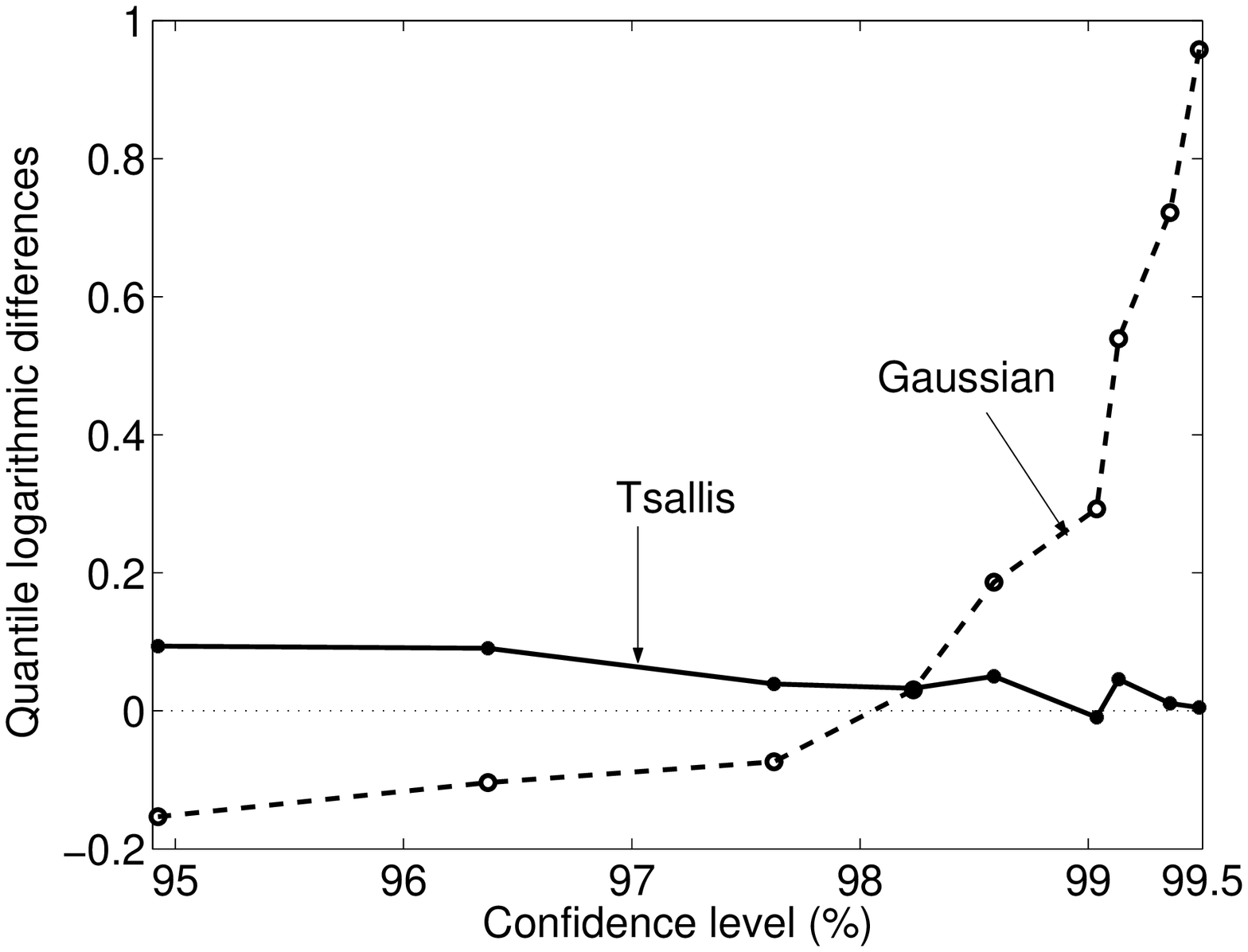,width=6.7cm,height=6.2cm}}
\caption{(a) Experimental ($\diamond$) and theoretical (Gaussian:
dashed line; Tsallis: solid line, $q=1.51$) cumulative density
functions (CDF); (b) Experimental and theoretical quantile
logarithmic differences as a function of the confidence level.}
\label{cdf_fig}
\end{center}
\end{figure}

Figure \ref{cdf_fig}a compares two theoretical cumulative density
functions (CDF), computed using Gaussian and Tsallis ($q=1.51$)
standardized return distributions, with CASI 1-day experimental
histograms. We observe that Tsallis distribution provides better
VaR estimates, for confidence levels within the range of 95 to
99.5\%. This comparison may be further enhanced plotting the
difference between experimental and theoretical quantiles as a
function of the confidence level (\ref{cdf_fig}b). Note that for
high confidence levels (above 99\%), the Gaussian model grossly
underestimates the risk, while for low confidence levels (below
97\%), it leads to sub-optimal capital allocation due to
systematic risk overestimation.

\section{Conclusion}

In this paper, we studied the financial risk of the aerospace
sector. In spite of its undeniable economic, social and even
strategic importance, until presently, there was no single market
indicator specially designed for tracking the performance of the
aerospace sector as a whole. This fact led us to develop a new
index, called Comprehensive Aerospace Index (CASI), based on the
stocks negotiated on the New York Exchange (NYSE) and the Over
the Counter (OTC) markets. Our data set consisted of twelve years
of daily prices sets, from January, 1987 to April, 1999, from a
comprehensive sample of leading aerospace-related companies.

We investigated the statistical characteristics of the CASI and
found that it is more volatile but less intermittent than other
traditional market indicators, such as the DJI and S\&P500. This
results suggests the existence of long memory correlations
impacting the volatility clustering patterns of the CASI index.
We also shown that CASI histograms, for different time intervals,
are well described by Tsallis canonical distribution. We explored
this result and modified the standard Value-at-Risk (VaR) risk
assessment methodology in order to reflect an asset whose
underlying dynamics obeys Tsallis non-extensive statistics. Our
results have shown this approach provides better VaR estimates,
for confidence levels within the range of 95 to 99.5\%.

Summarizing, we explored in this paper the idea that the
comprehension of financial time-series statistical features may be
the first step to grasp the behavior of a highly complex system
such the financial market. Investment professionals have at their
disposal a wide array of sophisticated tools to aid their
analyzes, like the Value-at-Risk method. Many of these tools
depend on the assumption that the asset returns are normally
distributed. Today we know they are not. To emphasize the
importance of this point, we showed in our results that the
relaxation of normality premise can improve a risk measure, with
positive impacts on the stability of markets and the
profitability of financial institutions.

\section*{Acknowledgments}
This work was supported by FAPESP, CAPES and CNPq, Brazil.

\end{document}